\documentclass[pre, twocolumn]{revtex4}
\usepackage{amssymb}

\newcommand{\be}{\begin{equation}}
\newcommand{\ee}{\end{equation}}
\usepackage[figuresright]{rotating}

\begin{document}

\title{Validation of Dunbar's number in Twitter conversations}
\author{Bruno Gon\c calves${}^{1,2}$}
\author{Nicola Perra${}^{1,3}$}
\author{Alessandro Vespignani${}^{1,2,4}$}

\affiliation{${}^{1}$Center for Complex Networks and Systems Research, School of Informatics and Computing, Indiana University, IN 47408, USA}
\affiliation{${}^{2}$Pervasive Technology Institute, Indiana University, IN 47404, USA}
\affiliation{${}^{3}$Linkalab, Complex Systems Computational Lab. - 09100 Cagliari Italy}
\affiliation{${}^{4}$Institute for Scientific Interchange, Turin 10133, Italy}

\begin{abstract}

Modern society's increasing dependency on online tools for both work and recreation opens up unique opportunities for the study of social interactions.
A large survey of online exchanges or conversations on Twitter, collected across six months involving $1.7$ 
million individuals is presented here. We test the theoretical cognitive limit on the number of stable social relationships known as Dunbar's number. We find that users can entertain a maximum of $100-200$ stable relationships in 
support for Dunbar's prediction. The ``economy of attention'' is limited 
in the online world by cognitive and biological
 constraints as predicted by Dunbar's theory. Inspired by this empirical evidence we propose a simple dynamical mechanism, based on finite priority queuing and time resources, that reproduces the observed social behavior.
\end{abstract}

\keywords{Computational Social Science \sep Dunbar Number \sep Twitter \sep Economics of Attention}
\pacs{}
\maketitle

\section{Introduction}
Modern society's increasing dependence on online tools for both work and recreation has generated an unprecedented amount of data regarding social behavior.
 While this dependence has made it possible to redefine the way we study social behavior, new online communication tools and media are also constantly redefining social acts and relations.
 Recently, the divide between the physical world and online social realities has been blurred by the new possibilities afforded by real-time communication and broadcasting, which appear
 to greatly enhance our social and cognitive capabilities in establishing and maintaining social relations. The combination of mobile devices with new tools like Twitter, Foursquare, Blippy,
 Tumblr, Yahoo! Meme, Google Hotspot, etc., are defining a new era in which we can be continuously connected with an ever-increasing number of individuals through constant digital communication composed
 of small messages and bits of information. Thus, while new data and computational approaches to social science \cite{lazer09-1, watts04-1, cho09-1} finally enable us to answer a large number of long-standing questions
 \cite{barabasi05-1,castellano09-1,gonzalez08-1}, we are
 also increasingly confronted with new questions related to the way social interaction and communication change in online social environments: What is the impact that modern technology has on social interaction?
 How do we manage the ever-increasing amount of information that demands our attention?
 In $1992$, R. I. M. Dunbar \cite{dunbar92-1} measured the correlation between neocortical volume and typical social group size in a wide range of primates and human communities.
 The result was as surprising as it was far-reaching. The limit imposed by neocortical processing capacity appears to define the number of individuals with whom it is possible
 to maintain stable interpersonal relationships. Therefore, the size of the brain's neocortex represents a biological constraint on social interaction that limits humans' social network
 size to between $100$ and $200$ individuals \cite{dunbar98-1}, i.e. Dunbar's number. McCarty et al. \cite{mccarty01-1} independently attempted to measure typical group size using two different methods and obtained a
 number of $291$, roughly twice Dunbar's estimate.\\
 Biological constraints on social interaction go along with other real-world physical limitations. After all, a persons time is finite and each person must make her own
 choices about how best to use it given the priority of personal preferences, interests, needs, etc. The idea that attention and time are scarce resources led H. Simon~\cite{simon71-1} to apply standard
 economic tools to study these constraints and introduce the concept of an Attention Economy with mechanisms similar to our everyday monetary economy. The increasingly fast pace of modern life
 and overwhelming availability of information has brought a renewed interest in the study of the economy of attention with important applications both in business \cite{davenport02-1} and the study collective
 human behavior \cite{huberman07-1}.  On the one hand it can be argued that microblogging tools facilitate the way we handle social interactions and that this results in an online world where human social
 limits are finally lifted, making predictions such as the Dunbar's number obsolete. Microblogging and online tools on the other hand, might be analogous to a pocket calculator that, while
 speeding up the way we can do simple math, does not improve our cognitive capabilities for mathematics. In this case, the basic cognitive limits to social interactions are not surpassed 
in the digital world. In this paper we show that the latter hypothesis is supported by the analysis of real world data that identify the presence of Dunbars limit in Twitter, one of the most successful
 online microblogging tools.\\

\begin{figure*}
\centering
    \includegraphics[width=16cm]{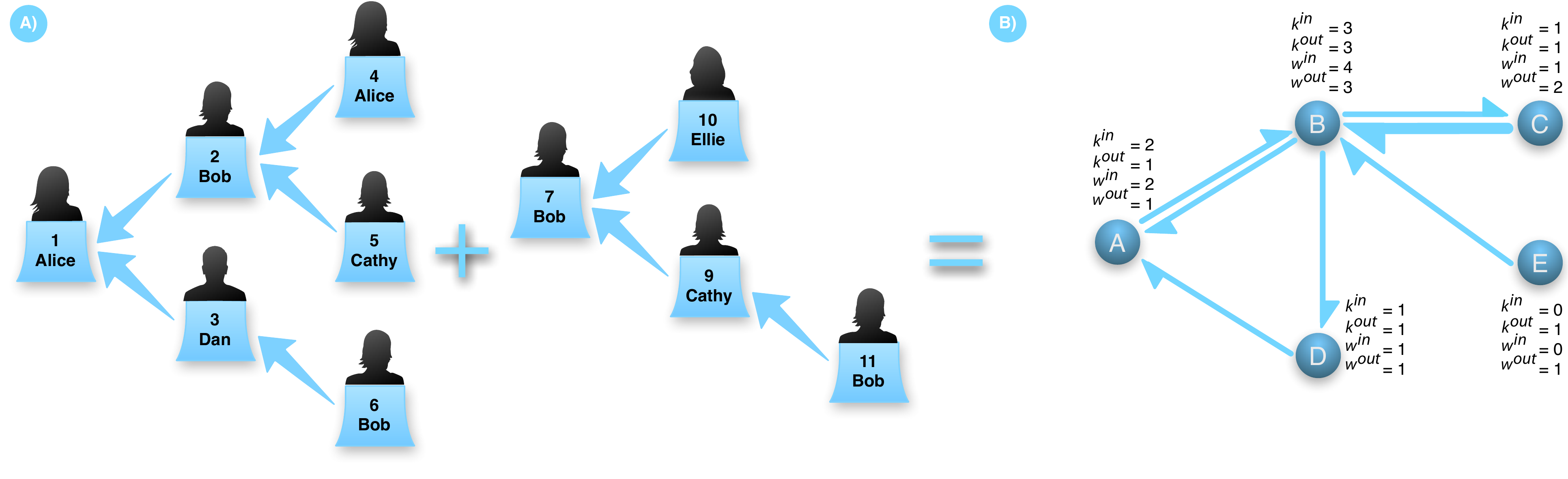}
    \caption{\label{fig1} Reply trees and user network. A) The set of all trees is a forest. Each time a user replies, the corresponding tweet is connected to another one,
 resulting in a tree structure. B) Combining all the trees in the forest and projecting them onto the users results in a directed and weighted network that
 can be used as a proxy for relationships between users. The number of outgoing (incoming) connections of a given user is called the out (in) degree and is represented by $k^{out}$ ($k^{in}$).
 The number of messages flowing along each edge is called the degree, $\omega$. The probability density function $P(k^{out})$ ($P(k^{in})$) indicates the probability that any given node has $k^{out}$ ($k^{in}$)
 out (in) degree and it is called the out (in) degree distribution and is a measure of node diversity on the network.}
\end{figure*}

\section{The Dataset}

Having been granted temporary access to Twitters firehose we mined the stream for over $6$ months to identify a large sample of active user accounts. Using the API,
 we then queried for the complete history of $3$ million users, resulting in a total of over $380$ million individual tweets covering almost $4$ years of user activity on
 Twitter. Table~\ref{stats} provides some basic statistics about our dataset.
Here we analyze this massive dataset of Twitter conversations accrued over the span of six months and investigate the possibility of deviation from Dunbar's number in the number
 of stable social relations mediated by this tool. The pervasive nature of Twitter, along with its widespread adoption by all layers of society, makes it an ideal proxy for the study
 of social interactions~\cite{huberman08-1,lashinsky08-1,boyd08-1,kwak10-1}.  We have analyzed over $380$ million tweets from which we were able to extract $25$ million conversations.
 Each Twitter conversation takes on the form of a tree of tweets,
 where each tweet comes as a reply to another. By projecting this forest of trees onto the users that author each tweet, we are able to generate a weighted social network connecting over $1.7$ million
 individuals (see Figure~\ref{fig1}).

\begin{table}[h!]
\centering
\begin{tabular}{lr} 
\hline\hline
Tweets & $381,652,990$\\
Timelined Users & $3,006,180$\\
Scraping Period & Nov. $20$, $2008$ -- May $29$, $2009$\\
Time span & $4$ years\\
\hline
\\
\hline
Trees & $25,273,871$\\
Tweets in Trees& $81,728,252$\\
User in Trees& $1,720,320$\\
User-User Edges & $68,459,592$\\
\hline\hline
\end{tabular}
\caption{\label{stats} Dataset Statistics.}
\end{table}

\subsection{Tree Identification and Projection} 

All tweets in our dataset that constituted a reply were collected. Each such tweet contains information not only about the id of the original
 tweet but also the user that sent it. Using this information, each reply tweet maps directly to a directed edge. Individual trees can be identified by using
 depth first search~\cite{cormen01-1} to identify connected components in the resulting tweet-tweet graph. To ensure that the full tree is found and not just a part of it, we
 treat each link as undirected for the purposes of this identification. In this way we are able to extract the complete tree even if we happen to start on one of the leaves. For
 each tree the root is then found by locating the node with $k_{in}\equiv 0$, and distances from the root are measured by rerunning the DFS algorithm starting from the root and respecting the direction of each edge.

The underlying reply network can be extracted by projecting the tweet trees to a user graph: User $A$ is connected to user $B$ by a directed outgoing edge if $A$ replied to a tweet sent by $B$.
 Over time, any pair of users can exchange multiple replies either in a single ``conversation'' (tree) or through multiple conversations. The number of messages sent from one user to another is
 used as the weight of the corresponding directed edge and is taken to signify the strength of the connection between the two users, with higher weights representing stronger connections.

\begin{figure*}
\begin{center}
\begin{tabular}{cc}
\includegraphics[width=8cm]{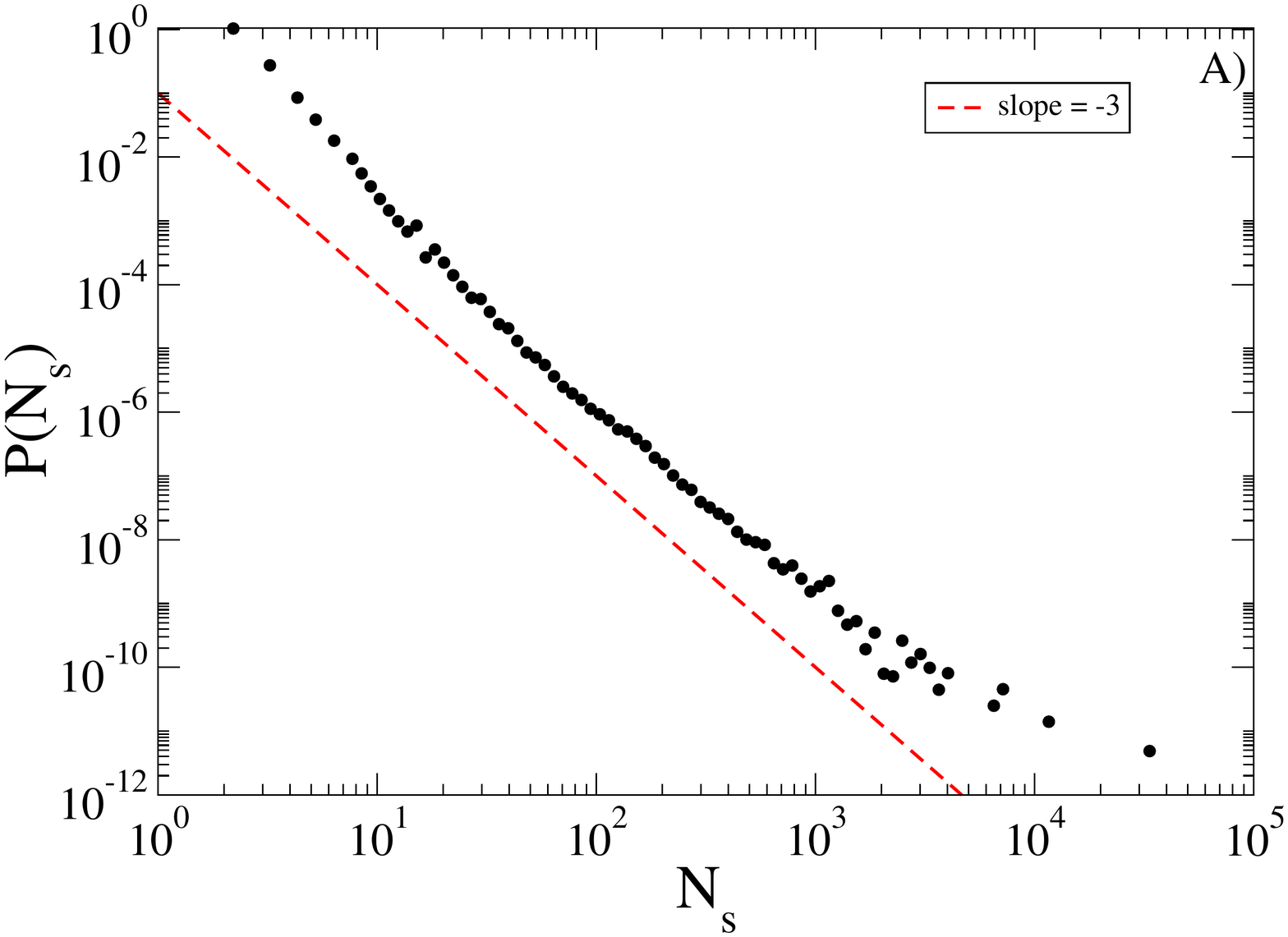}&\includegraphics[width=8cm]{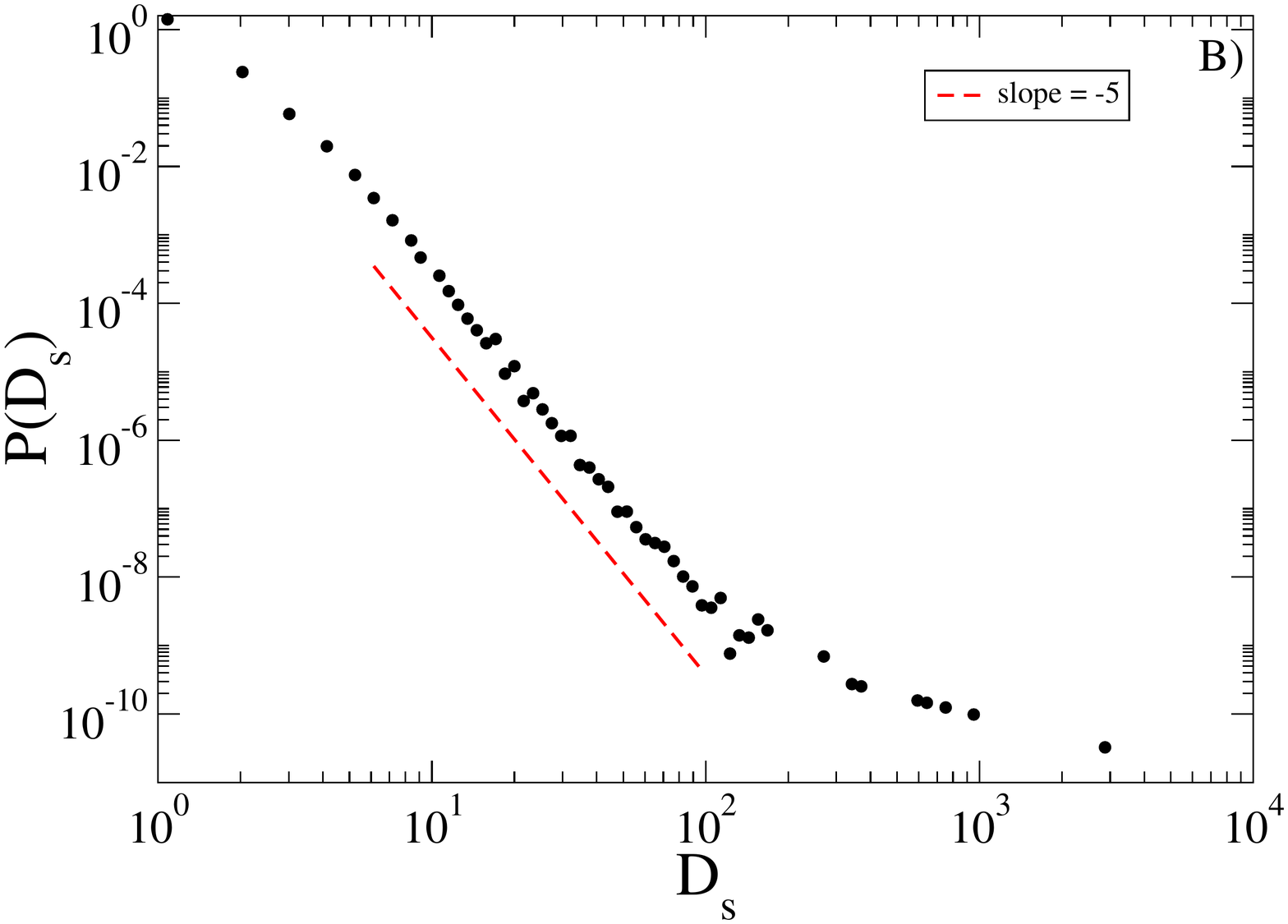}\\
\includegraphics[width=8cm]{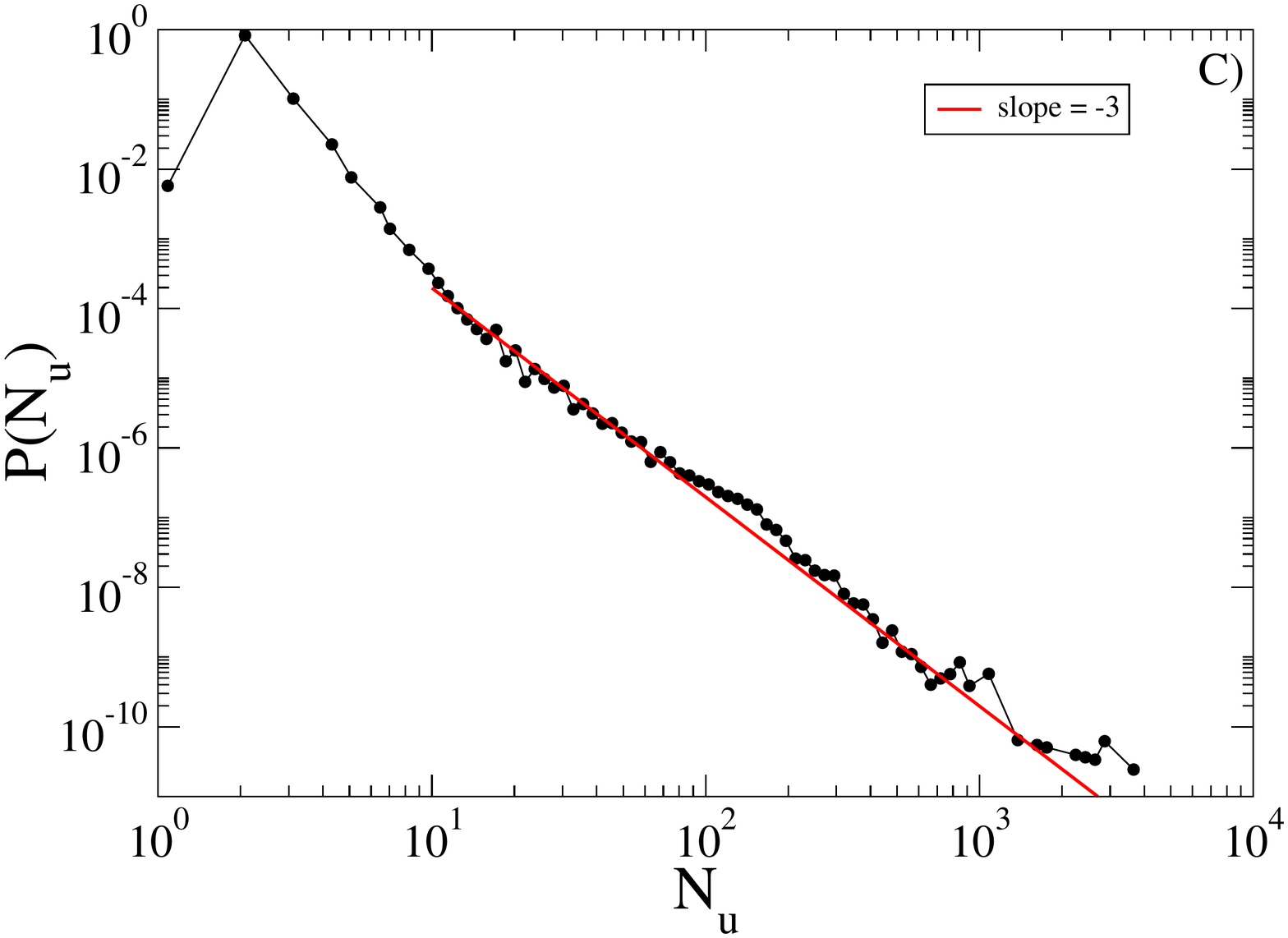}&\includegraphics[width=8cm]{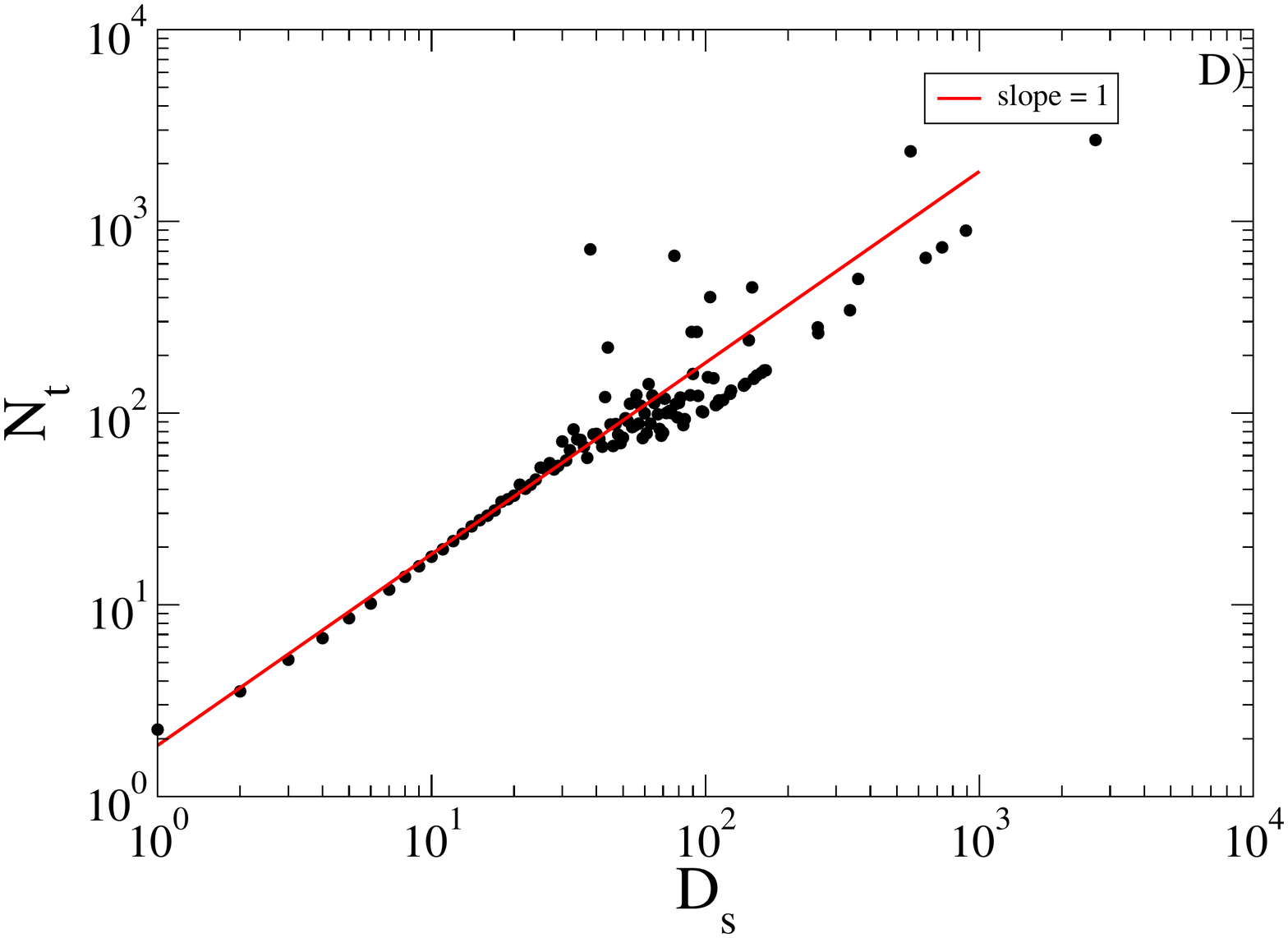}\\
\end{tabular}
\end{center}
\caption{\label{Tree-characterization}Tree characterization. A)Distribution of the number of tweets in a tree. B) Distribution of the number of shells. C) Distribution of the number of users. D) Tree size vs depth. The broad tailed nature of all of these quantities indicates the diversity of behaviors displayed by the users in our dataset.}
\end{figure*}

\subsection{Online conversations}

Each reply creates a connection between two tweets and their authors, so we can define a conversation as a branching process of consecutive replies, resulting in a tree of tweets.
 From our dataset we extracted and analyzed a forest of over $25$ million trees. Trees vary broadly in size and shape, with most conversations remaining small while a few grow to include 
thousands of tweets and hundreds of users, as shown in Figure~\ref{Tree-characterization}. 

A directed user-user network can be built by projecting conversation trees to detail how users interact and establish relationships among themselves. Bidirectional edges signify mutual
 interactions, with stronger weights implying a more frequent or prolonged interaction between two individuals.

All of our analysis will be performed on this user-user conversation network. We consider a user to have out degree $k_{out}$ if he or she replies to $k_{out}$ other users, regardless of
 the number of explicit followers or friends the given user has. By focusing on direct interactions we are able to eliminate the confounding effect of users that have tens or hundreds of
 thousands of followers with whom they have no contact and are able to focus on real person to person interactions~\cite{huberman08-1}.
\section{Dunbar's number in our data}
 
In the generated network each node corresponds to a single user. The out-degree of the nodes is the number of users the node replies to, while the in-degree corresponds to the number
 of different nodes it receives a reply from. When A follows B, A subscribes to receive all the updates published by B. A is then one of B's followers and B is one of A's friends.
 Previous studies have mostly focused on the network induced by this follower-friend relationship~\cite{boyd08-1}, \cite{krishnamurthy08-1,java07-1,cha10-1}. In any study about stable social relations in online media, as indicated by
 studies about Dunbar's number, it is important to discount occasional social interactions. For this reason we focus on stronger relationships  in our study\cite{huberman08-1}, considering just active communication
 from one user to another by means of a genuine social interaction between them. In our network~\cite{barrat08-1,newman10-1}  we introduce the weight $\omega_{ij}$ of each edge, defined as the number of times user $i$ replies to user $j$
 as a direct measurement of the interaction strength between two users and stable relations will be those with a large weight. A simple way to measure this effect is to calculate the average weight of each
 interaction by a user as a function of his total number of interactions. Users that have only recently joined Twitter will have few friends and very few interactions with them. As time goes by, stable users
 will acquire more and more friends, but the number of replies that they send to other users will increase consistently only in stable social interactions. Eventually, a point is reached where the number of
 contacts surpasses the user's ability to keep in contact with them.\\ 
This saturation process will necessarily lead to some relationships
being more valued than others. Each individual tries to optimize her
resources by prioritizing these interactions. To quantify the strength
of these interactions, we studied the quantity $\omega_i^{out}$ , defined as the
average social strength of active initiate relationship:
\be
\label{omega}
\omega_{i}^{out}(T)=\frac{\sum_{j}\omega_{ij}(T)}{k_i^{out}}
\ee

\begin{figure*}
\centering
    \includegraphics[width=16cm]{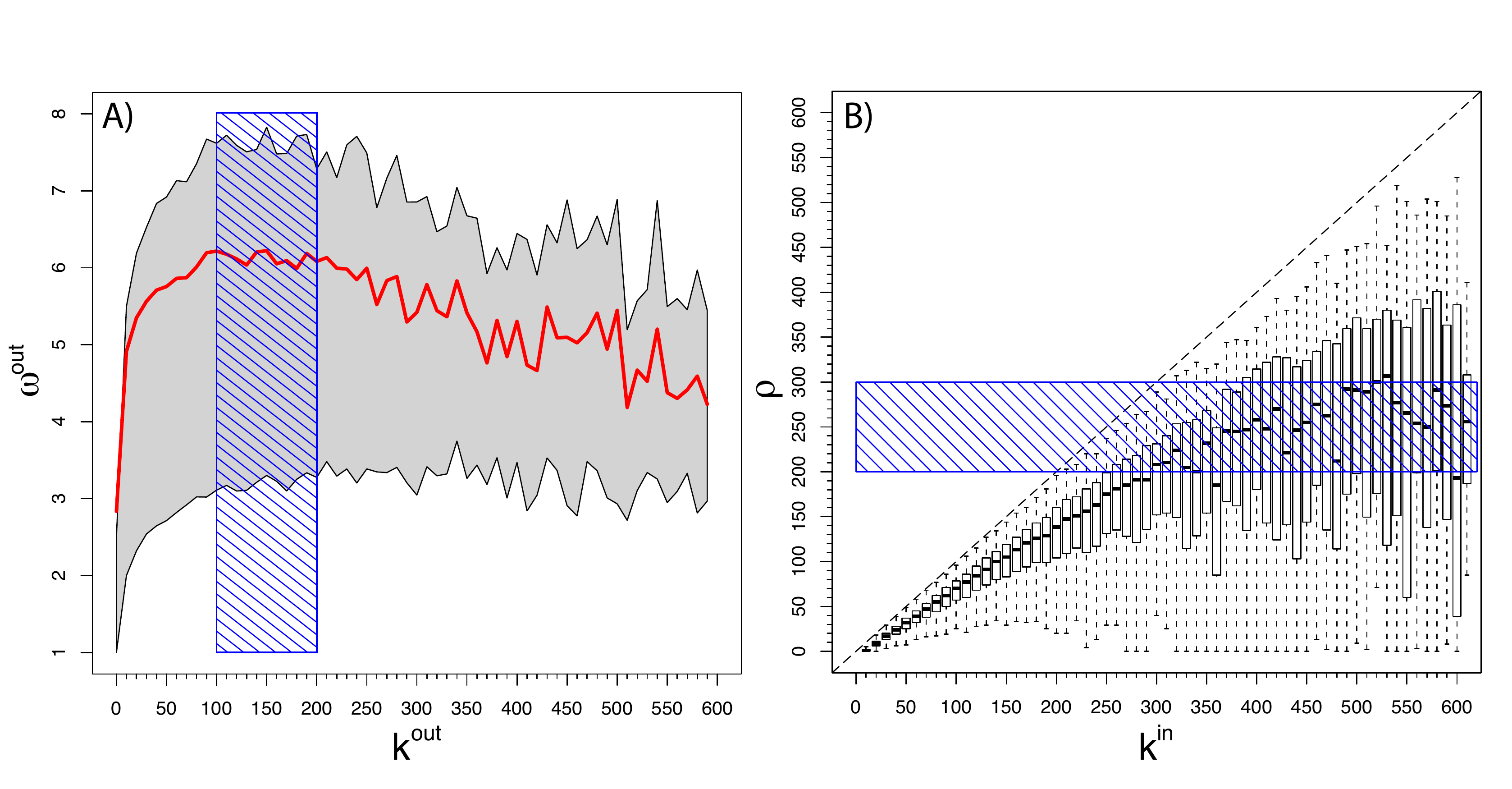}
    \caption{\label{fig2}  A) Out-weight as a function of the out-degree. The average weight of each outward connection gradually increases until it reaches a maximum near $150-200$
 contacts, signaling that a maximum level of social activity has been reached. Above this point, an increase in the number of contacts can no longer be sustained with the same amount of dedication to each.
 The red line corresponds to the average out-weight, while the gray shaded area illustrates the $50\%$ confidence interval. B) Number of reciprocated connections, $\rho$, 
as a function of $k_{in}$. As the number of people demanding our attention increases, it will eventually saturate our ability to reply leading to the flat behavior displayed in the dashed region.}
\end{figure*}

This quantity corresponds to the average weight per outgoing edge of each individual where $T$ represents the time window for data aggregation. We measure this quantity in our data set as shown in Figure~\ref{fig2}A.
 The data shows that this quantity reaches a maximum between 100 and 200 friends, in agreement with Dunbar's prediction (see figure 2A). This finding suggests that even though modern social networks help us
 to log all the people with whom we meet and interact, they are unable to overcome the biological and physical constraints that limit stable social relations. In Figure 2B, we plot , the number of reciprocated 
connections, as a function of the number of the in-degree.  saturates between 200 and 300 even though the number of incoming connections continues to increase. This saturation indicates that after this point
 the system is in a new regime; new connections can be reciprocated, but at a much smaller rate than before. This can be accounted for by spurious exchanges we make with some contacts with whom we do not maintain
 an active relationship.
 
\section{The model}

\begin{figure}
\centering
    \includegraphics[width=8cm]{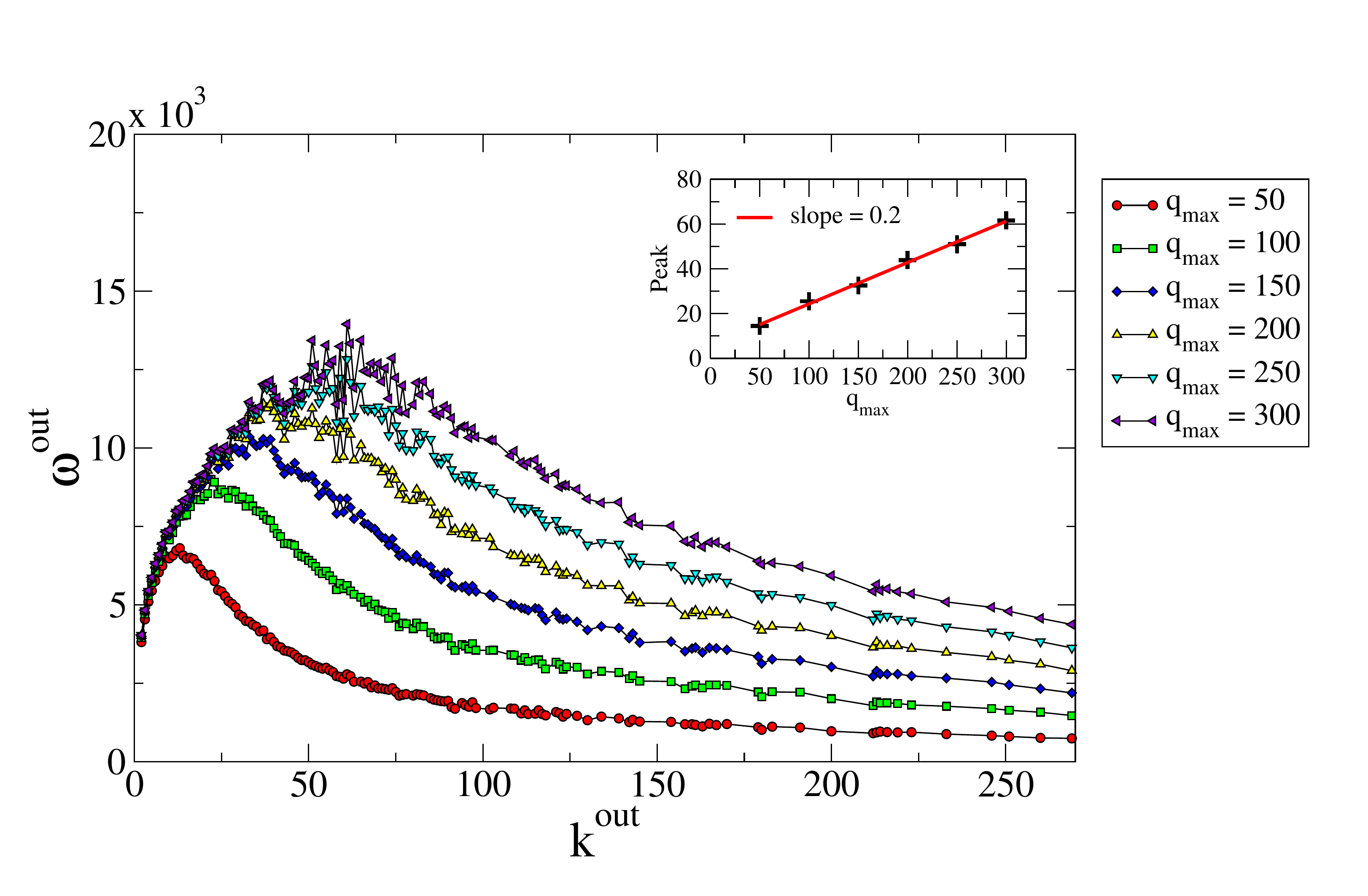}
    \caption{\label{fig3} Result of running our model on a heterogeneous network made of $N=10^5$, nodes with degree distribution $P(k)~k^{-\gamma}$ with $\gamma=-2.4$ and $\sigma=10$.
 Different curves correspond to different queue size. The inset shows the linear dependence of the peak on the queue size $q$.  Each curve is the median of $10^3$ to $2\times 10^3$ runs of $T=2\times 10^4$ time steps}
\end{figure}

Let us consider a static network $\mathcal{G}$, characterized by a
degree distribution $P(k)$. Each user
(node) $i$ is connected to all its nearest neighbors $j$s through two weighted directed edges, $i\to j$ and $j\to i$ so that:
\be
k_i^{out}=k_i^{in}=\frac{k_i}{2}, \;\ \forall i \in \mathcal{G}.
\ee
Where $k_i^{out}$ is the out degree, the number of out-going
links, and
$k_i^{in}$ is the in degree, the number of in-going links, of the user
$i$. Each node uses its out links to send messages to its contacts and
it will receive messages from its contacts through its in links. In this way is easy to distinguish between incoming and
outgoing messages. Whenever a message is sent from node $i$ to node $j$,
the weight of the $(i, j)$ edge, $w_{ij}$ is increased by one. The
total number of sent messages of each user is given by the sum over all of its
outgoing edges. Users communicate with each other by replying to
messages.  The assumption of our model is that biological and time constraints are the keys ingredients in fixing the Dunbar's number. We model this considering that when user $i$ receives a message it places it in an
internal queue that allows up to $q_{max,i}$ messages to be handled at each
 time step. In the presence of finite resources each agent has
 to make decisions on what are the most important messages to
 answer. We set the priority of each message to be proportional to the total degree of the sender $j$. For each user
 the we studied is the average number of interactions per
 connection $\omega_i^{out}(T)$ as defined in the Eq.~(\ref{omega}).
At each time step each agent goes through its queue and performs the
  following simple operations:
\begin{itemize}
\item       The agent replies to a random number $S_t$ of messages between 0 and the number of messages $q_{i}$ present in the queue. The messages to be replied to are selected proportionally to 
the priority of the sending agent (its total degree).  A message is then sent to $j$, the node we are replying to, and the corresponding weight $\omega_{ij}$ is incremented by one.
\item       Messages the agent has replied to are deleted from the queue and all incoming messages are added to the queue in a prioritized order until the number of messages
 reaches $q_{max}$. Messages in excess of $q_{max}$ are discarded.
\end{itemize}
The dynamic process is then repeated for a total number of time steps $T$. In order to initialize the process and take into account the effect of endogenous random effects, each agent can
 broadcast a message to all of its contacts with some small probability $p$. One may think of this message as a common status change, or a TV appearance, news story, or any other information
 not necessarily authored by the sending agent. Since these messages are not specifically directed from one user to another, they do not contribute to the weight of the edges through which they
 flow. We have studied this simple model by using an underlying network of $N=10^5$ nodes and different scale-free topologies. For each simulation $T=2\times10^4$ time steps have been considered and the plots
 are made evaluating the medians among at least $10^3$ runs.
In Figure~\ref{fig3} we report the results of simulations in a directed heavy-tailed network with a power-law tail similar to those observed for the measured network~\cite{java07-1}. The figures
 clearly show a behavior compatible with the empirical data. The peak that maximizes the information output per connection is linearly proportional to $q_{max}$, supporting the idea that the physical constraints
 entailed in the queue's maximum capacity along with the prioritization that gives importance to popular senders are at the origin of the observed behavior. We have also performed an extensive sensitivity
 analysis on the broadcasting probability $p$, the time scale $T$, and have investigated the effect of agent heterogeneity by studying populations in where each agent's capacity $q_{max}$,$i$ is randomly distributed
 according to a Gaussian distribution centered around $q_{max}$ with standard deviation $\sigma$.
\subsection{Effect of the time window $T$}

One of the parameters of our model is the time window $T$ during which we study the dynamics. This parameter regulates the maximum number of messages that will circulate in the network. In the first time steps 
the first messages will start to being sent among users and the queues start to get messages in and out. After a while we can aspect that the system reaches a dynamical equilibrium. In Figure~(\ref{unique}-A) we show 
the behavior of our observable $\omega^{out}$ for different values of $T$, in particular we chose $T=10^4, 1.5\times 10^4, 2\times 10^4$. The effect of time is clearly a shift on the y axis and a small change in the
 position of the peak. The first effect is due to the fact that the number of messages circulating in the systems increase linearly with $T$. The second effect is due to the reduction of fluctuations when more 
messages are sent. The peak becomes more clear and defined.
\subsection{Effect of broadcast probability $p$}

The effect of the broadcast probability is different on respect to the effect of the time window $T$. First of all our observable $\omega^{out}$ is linearly proportional to $T$ in all regimes of $k^{out}$ this 
is not true for $p$. The effect of $p$ is crucial for users with a small number of contacts. As the $p$ increases they will receive more messages and their activity will increase too, this does not occur 
in the other limit. When the saturation takes place the $\omega^{out}$ becomes completely independent of $p$. 
 As show in details in a mean-field approach (Section~(\ref{singleguy})) for values of $k^{out}$ small with respect to the queue size, $\omega^{out}$ scales linearly with $p$. Instead for a number of contacts 
much bigger than the queue size $\omega^{out}$ is independent of $p$. These considerations are validated by our simulations as shown in Figure~(\ref{unique}-B). We see a clear dependence on $p$ for small values of 
$k^{out}$ instead the same behavior for bigger values of $k^{out}$.
\subsection{Effect of network's properties}

\begin{figure*}
\centering
    \includegraphics[width=16cm]{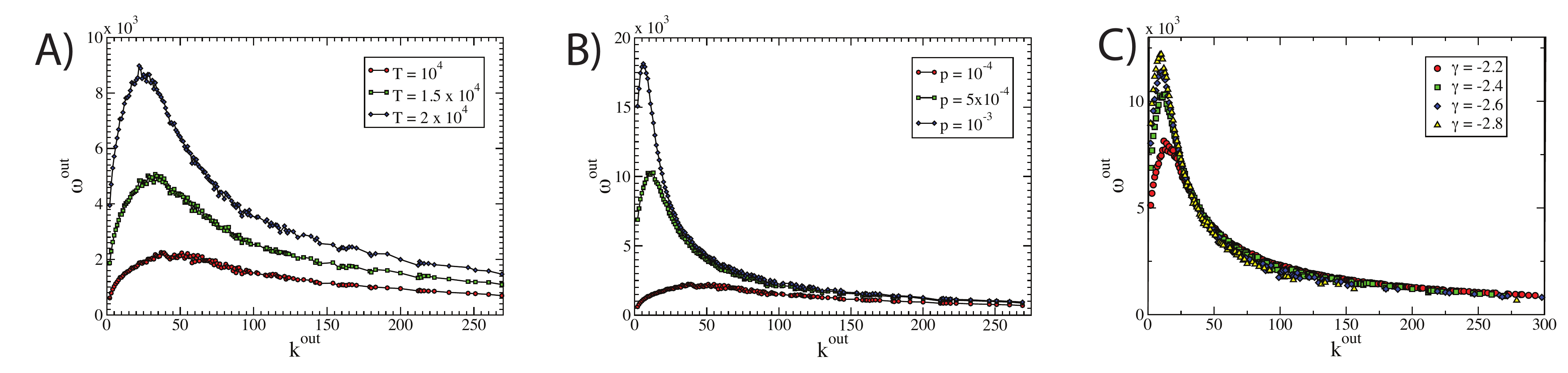}
    \caption{\label{unique} A) $\omega^{out}$ as a function of
      $k^{out}$ for $q_{max}=100$, $\sigma=10$, scale-free network
      with $\gamma=-2.4$ and different values of $T$. We present the
      medians over $500$ runs B)
$\omega^{out}$ as a function of $k^{out}$ for $q_{max}=100$, $\sigma=10$, scale-free network with $\gamma=-2.4$, $T=10^4$ and different values of $p$. We present the medians over $500$ runs C) $\omega^{out}$ as a function of $k^{out}$ for $q_{max}=100$, $\sigma=10$, $T=10^4$, $p=5\times 10^{-4}$, scale-free network with different values of $\gamma$. We present the median over $500$ runs.}
\end{figure*}

Inspired by several studies~\cite{java07-1,boyd08-1,huberman08-1} we fix the baseline of our model using scale-free networks. It is important then to study how differences in the network structure affect the results. 
In this section we consider the effect of the 
exponent $\gamma$. As show in Figure~(\ref{unique}-C) we run our model on top of scale-free networks with $\gamma=-2.2,-2.4, -2.6, -2.8$. As clear from the plot for smaller values of $\gamma$
 (bigger value in absolute value)
 gaps on $k^{out}$ start to emerge. These are due to the network structure. The shape and position of the peak is the same for all the curves. The differences are evident just on the 
peak height that increase as $\gamma$ decreases. This is due the different redistribution of degrees and to the fact that with small $\gamma$ the selection effect is more and more important. So we can say 
that the result are robust on $\gamma$. 
\subsection{Single user: analytical approach}
\label{singleguy}

In order to get a better understanding of the mechanisms we describe,
we analyzed, in a mean-field approach, the behavior of a single user $i$.\\
 Let us focus on a user $i$ characterized by degree $k_i$ and
 $q_{max,i}$. $k_i^{out}=k_i/2$ are the out-going links that it uses
 to send messages to its $k_i/2$ contacts. $k_i^{in}=k_i/2$ are the
 in-coming through which it receives messages from it contacts. We set
 as $k_j$ the priority of each neighbor  that we extract for a distribution $\mathcal{P}(k)$.
 The rules of the model that we described in the previous sections
are applied for $T$ time steps. The probability that a neighbor $j$ will send a message to the user $i$ is:
\be
p_{ji}=p+\frac{k_i}{<k>k_j},
\ee  
where $p$ is the broadcast probability. We can evaluate the average number of messages that the user will receive at each time step $t$:
\be
\langle R\rangle=\sum_{j}p_{ji}=pk_i^{in}+\frac{k_i}{<k>}\sum_{j \in k_i^{in}}\frac{1}{k_j}.
\ee
We extracted $k_j$ from the same distribution, the sum scales then
linearly with the number of element: $k_i^{in}$. We can write:
\be
\label{m_t}
\langle R\rangle=\sum_{j}p_{ji}=p\frac{k_i}{2}+c\frac{k_i^{2}}{2<k>},
\ee
where $c$ is a constant fixed by the distribution. Since the priority
of the user is proportional to its degree as well as the number 
of in-coming connections, the number of messages it get scale as the square of its degree.\\
Two different regimes are easily found: $k_i \ll q_{max,i}$ and vice versa.\\
In the first case the user is not popular. The number of messages that
the user will receive is small then. In principle it can reply to all of them at each time step. We can assume that in this regime 
its queue is never completely full.  We will refer to $R_t$ as the
number of messages that the user reiceive at the time step $t$. After one time step the number of replies is:
\be
S_1=\xi_1 R_1,
\ee 
where $\xi_1$ is a random number uniformly distributed between $0$
and $1$. The number of messages, $S_2$ that the user send at the second time step is a random fraction of the messages present in its queue:
\be
S_2= \left [ R_1(1-\xi_1)+R_2 \right]\xi_2.
\ee
For $t=3$ we get:
\begin{eqnarray}
S_3&=& \left\{\left[ R_1(1-\xi_1)+R_2 \right](1-\xi_2)+R_3 \right \}\xi_3\nonumber\\
&=& \left[ R_1(1-\xi_1)(1-\xi_2)+R_2(1-\xi_2)+R_3 \right]\xi_3,
\end{eqnarray}
and so on. We can approximate these equations using  the average
number of received messages $\langle R\rangle$. For the general $t$ it
is possible to show that:  
\begin{eqnarray}
S_t&\sim& \langle R\rangle\xi_t \left[1+ \sum_{j=1}^{t}\prod_{i=j}^{t-1}(1-\xi_i) \right] \nonumber \\
&=& \langle R\rangle\xi_t \left[2- \xi_{t-1}+\mathcal{O}(\xi^2)\right]\nonumber\\
&=& \langle R\rangle \left[ 2\xi_t-\xi_t\xi_{t-1}+\mathcal{O}(\xi^3) \right].
\end{eqnarray}
The total number of messages sent is the numerator of our measure $\omega^{out}$ and the sum of all the $S_t$:
\be
\sum_{t=0}^{t=T}S_t\sim T\langle R\rangle,
\ee
considering that each sum of product random numbers is order $T$. We can write then:
\begin{eqnarray}
\omega^{out}_{i}(T)&=&\frac{\sum_{t=0}^{t=T}S_t}{k_i^{out}} \sim \frac{1}{k_i^{out}}T \left[ p\frac{k_i^{out}}{2}+c\frac{(k_i^{out})^2}{2<k>} \right ]\nonumber\\
&\sim& T\,\ k_i^{out}.
\end{eqnarray}
\begin{figure}
\centering
    \includegraphics[width=8cm]{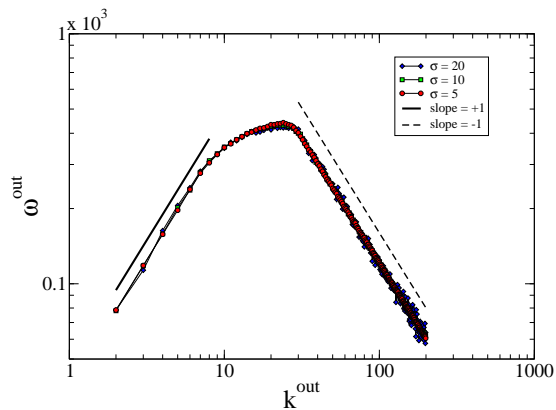}
    \caption{\label{fig4} Results for the single user and different values of $\sigma$, the inter-user queue size variance. We fixed the average queue size at $q_{max,i}=50$ and extracted the priorities
 of user neighbors from a power-law statistical distribution with exponent $\gamma=-2.1$. For each $k_i$ we run $T=500$ time steps and present the medians among $10^3$ runs}
\end{figure}

In this regime we get a linear increase with $k_i^{out}$ of the average number of replies per connections. As show in Figure~(\ref{fig4}) this is confirmed in the simulations.\\
The other regime is found for a number of contacts bigger than the
queue size. In this case the user is very popular and at each time
step it gets a lot of messages and is not able to handle all of it. In
this limit the saturation 
process takes place and it will reply just to a small fraction of the total number of messages prioritizing them. At each time step this number is a random variable uniformly distributed
 between $0$ and $q_{max,i}$. We have then:
\be
\omega^{out}_{i}(T) \sim \frac{1}{k_{i}^{out}}\sum_{t=0}^{T} \xi_t q_{max,i}.
\ee 
The $\xi_t$s are random variable uniformly distributed between $0$ and
$1$. At each time step the number of replies is a random fraction of the queue size. For $T$ large enough we get:
\be
\omega^{out}_{i}(T)\sim \frac{T}{2k_{i}^{out}}q_{max,i}.
\ee
In this regime then we get a different scaling behavior typical of
saturation problems. As shown in Figure~(\ref{fig4}) these arguments
are in perfect agreement with the numerical results.\\
We have shown two different regimes. A linear increasing behavior and a decreasing one. In the between of these opposite cases we will find a maximum of the function. The position of these peak is in general function 
of the queue size.
\section{Conclusions}

Social networks have changed they way we use to communicate. It is now easy to be connected with a huge number of other individuals. In this paper we show that social networks did not change 
human social capabilities. We analyze a large dataset of Twitter conversations collected across six months involving
millions of individuals to test the theoretical cognitive limit on the number of stable social relationships known as Dunbar's number.  We found that even
in the online world cognitive and biological
 constraints holds as predicted by Dunbar's theory limiting users social activities. We propose a simple model for users' behavior that includes finite priority queuing and time resources that reproduces the observed social behavior. This simple model offers a basic explanation of
a seemingly complex phenomena observed in the empirical patterns on
Twitter data and offers support to Dunbar's 
hypothesis of a biological limit to the number of relationships. 

\section{Acknowledgements}

The work has been partly sponsored by the Army Research Laboratory and
was accomplished under Cooperative Agreement Number W911NF-09-2-0053.
The views and conclusions contained in this document are those of the
authors and should not be interpreted as representing the official
policies, either expressed or implied, of the Army Research Laboratory
or the U.S. Government.

\end{document}